\documentclass[preprint,12pt,authoryear]{elsarticle}

\usepackage{amssymb}
\usepackage{amsmath}
\usepackage{booktabs}
\usepackage[dvipsnames]{xcolor}

\journal{Coastal Engineering}

\begin{document}

\begin{frontmatter}



\title{Real-scale Smoothed Particle Hydrodynamics Tsunami Runup Modelling, with application to 3-D tsunami urban flows in Cilacap, South Java, Indonesia} 

\author[ucl]{Jack Dignan \corref{cor}}
\author[edi]{Joseph O'Connor}
\author[ucl]{Serge Guillas}

\affiliation[ucl]{organization={Department of Statistical Science},
            addressline={University College London}, 
            city={London}, 
            country={United Kingdom}}

\cortext[cor]{Corresponding author: Jack Dignan, jack.dignan@ucl.ac.uk}

\affiliation[edi]{organization={EPCC},
            addressline={University of Edinburgh}, 
            city={Edinburgh}, 
            country={United Kingdom}}

\begin{abstract}
The risk posed by tsunami waves is currently modelled over bare-earth representations by tsunami models. The complex flows around buildings and structures are crucial to represent the true state of the tsunami wave elevation, speed and forces exerted on buildings. Such 3D simulations have been unachievable for real scale modelling at a reasonable computational cost. We present here for the first time the use of Smoothed Particle Hydrodynamics (SPH) for tsunami simulation in a real setting of large scale (around 1 km). Our illustration is for Cilacap, Indonesia as constitutes a blueprint for future scenarios in South Java. SPH allows for the efficient modelling of shocks and complex interations of the flows with the structures. We also offer a range of test cases of increasing complexity and realism to tune and validate such realistic simulations, including the well known simplified beach of Seaside, Oregon at scale 1:50. We are able to reproduce realistic wave heights, velocities and even observed eddies. We provide guidance and discuss the various choices in terms of flow parametrisations, boundary conditions, and the trade-off of fidelity. computational cost. As a result, Probabilistic Tsunami Risk Assessments (PTRA) will become possible by making use a of combination of regional modelling of tsunamis with depth-average models (generation and propagation) as well as coastal modelling using SPH. 

\vspace{0.1cm}
\end{abstract}




\end{frontmatter}

%
%

\section{Introduction}

Tsunami waves present a significant threat to the lives, livelihoods, and economies of coastal communities. A key factor in improving evacuation planning, performing accurate impact assessments, and protecting critical infrastructure is to improve our understanding of the interaction between coastal structures and tsunami waves. The flow of fluid flows around obstacles and simplified building arrays has been widely studied both experimentally and numerically \citep{ishii2021experimental,wuthrich2020forces}. However, the simulation of tsunamis inundating urban scale environments to conduct probabilistic hazard analysis for real-world towns and cities remains largely unexplored. 

Existing tsunami numerical modelling that form the basis for the vast majority of probabilistic tsunami hazard analysis (PTHA) uses bare earth topography and roughness coefficients to represent urban topography, yet this will fail to represent the more local effects of real-world urban topographies on the overland propagation of tsunamis. Work conducted by \citet{moris2021tsunami} showed that rows of buildings provide a clear sheltering effect of tsunami overland flow, and provide clear differences in maximum inundation levels, cross-shore velocity, and cross-shore momentum flux values when compared to bare-earth simulations. This work also demonstrated that overland propagation was very sensitive to building offset and spacing, demonstrating the need to simulate real-world building arrays beyond gridded building geometries.

However, capturing these complex flows around buildings and structures requires a numerical approach capable of resolving turbulent flows across a high-resolution urban environment. One option is to solve the 3D Navier–Stokes equations using a traditional mesh-based CFD approach, see \citet{marras2020modeling} and references therein. However, such approaches face severe challenges in the presence of highly-deforming interfaces and violent free-surface flows. Therefore, in this work, we present the use of Smoothed Particle Hydrodynamics (SPH) for tsunami simulation through a range of test cases of increasing complexity and realism. Indeed, SPH outperforms traditional mesh-based for these types of flows thanks to its Lagrangian formulation, making it well suited to capturing the complex dynamics exhibited by tsunami inundation \citep{Violeau2016a}. The importance of this work is that it presents the first step in using SPH as a reliable numerical method for future PTHA. It is critical that we reach a level of confidence in the numerical model -- this includes a thorough understanding of SPH parameter sensitivity -- so that its outputs are meaningful and trustworthy. 

We begin by validating SPH against an experimental case setup of tsunami flow around a simple breakwater within a wavetank of uniform depth. Further experimental validation is shown with the second case representing a solitary wave runup on a simplified beach of Seaside, Oregon, which consists of a simplified building array atop synthetic bathymetry and topography. We validate both of these cases against experimental datasets to observe convergence and calibrate some of the more influential numerical parameters. Using these calibrated parameters, we then continue this development into creating an SPH model using real-world topography, bathymetry, and buildings for a site in the city of Cilacap, Central Java.

\section{Numerical Model}

The SPH method has been trialled in both academic and commercial settings to model fluid dynamics. Its use has been tested in marine engineering in vessel design \citep{tagliafierro2021performance,kawamura2016sph}, in offshore floating platform design \citep{dominguez2019sph}, in wave energy converter modelling \citep{marrone2019extreme}, and the simulation of sea waves impacting piers \citep{wei2015sph}. The use case of using SPH, and particularly DualSPHysics, for coastal engineering applications is one that has captured the attention of a relatively small group of researchers who continue to publish on the variety of applications for which SPH can be reliably used.

DualSPHysics has been used in coastal engineering to simulate wave impact on coastal defences, particularly coastal breakwaters \citep{zhang2018dualsphysics,liu2020numerical,armono2021numerical} and the research and development process of coastal defences \citep{yamamoto2021numerical,ma2021study}. More importantly, this has developed into modelling increasingly realistic coastal engineering cases. For example, the use of real world surfaces where \citet{barreiro2014integration} made use of UAV photogrammetry to recreate real world land surfaces to simulate overland flows from runoff events using DualSPHysics. In \citet{altomare2020sph}, we see the use of real world waves applied to a simplified pier structure along idealised bathymetry, producing similar wave profiles and measured impacts as was captured in the experimental replica of the case setup. Some effort has been made to simulate extreme wave flows across building arrays but with simplified or planar beach slopes \citep{rajapriyadharshini2022tsunami}, or using simplified input waves or buildings that do not accurately capture the large wavelength and mass of the fluid characteristic of tsunami flows \citep{klapp2020tsunami}.

Our work is providing a meaningful progression towards the use of SPH in coastal engineering with the use of real-world data. We aim to fill this knowledge gap, after validating the use of SPH in other simplistic cases that have experimental data to validate against, using real bathymetry and topography, with real world structures, and waveforms that are adequately resolved in the context of tsunami flows.

There is significant novelty in combining all of these elements which have been independently modelled in published literature to allow us to take a step in advancing the insights that tsunami simulations can provide. As documented in this work, the SPH method is able to compute forces exerted on structures, velocities through channels and around obstacles, and capture free surface elevations in turbulent flows with high levels of accuracy and configurability. Being able to extract these values across real-world coastal cities will present a significant advancement in tsunami preparedness and has the potential to act as the first step in making use of state-of-the-art methodologies and technologies to explore beyond traditional constraints of what is used around the world to prepare and inform coastal communities of estimated impact of tsunami events when they occur.

\subsection{SPH Formulation} \label{sec:sphformulation}
We present here a summary of the SPH solver and boundary conditions implemented within DualSPHysics. As this work does not seek to develop the numerical method or source code of DualSPHysics, we only aim to provide context to the parameter calibration conducted in this paper. For a more comprehensive perspective into the SPH formulation in DualSPHysics, see \citet{dominguez2022dualsphysics}.

\subsubsection{Particles and Interpolation}
\vspace{0.15cm}
In DualSPHysics, the fluid domain is divided into discrete particles, each representing a small part of the fluid. These particles do not represent actual particles, but instead are a method of discretising the continuous fluid domain. Each of these particles stores physical properties such as location ($x$ and $y$), velocity ($v$), pressure ($P$), and density ($\rho$), among others. However, to represent and manipulate quantities of interest that are continuous, SPH employs a smoothing technique, as follows.

The continuous approximation of a discontinuous function, $f(r)$, in a domain, $\Omega$, can be approximated by convolving $f$ with a kernel function, $W$:

\begin{equation}
    \langle f(\textbf{r}) \rangle = \int_{\Omega} f (\textbf{r'}) W (\textbf{r} - \textbf{r'}, h) d \textbf{r'}
\end{equation}

This convolution integrates the function $f$ over the domain, $\Omega$, weighted by the kernel $W(\textbf{r} - \textbf{r'}, h)$, where $h$ is a smoothing length that determines the scale over which the averaging occurs. 

The kernel function, $W(\textbf{r},h)$ is generally expressed as

\begin{equation}
    W(\textbf{r},h) := \frac{1}{h^d} \omega (q),
\end{equation}

\noindent where $d$ is the number of spatial dimensions (2 or 3), $q=\frac{|r|}{h}$ is the normalised distance, and $\omega(q)$ is a smooth, non-negative function that typically has compact support, meaning that it vanishes for distances beyond a certain multiple of $h$, represented by $k$:

\begin{equation}
    \omega (q) = 0 \text{ for } |r| \geq kh
\end{equation}

The kernel function has the important property of normalisation, ensuring the preservation of mass across the fluid domain:

\begin{equation}
    \int_{\Omega} W (\textbf{r},h)dr = 1,
\end{equation}

\noindent which, given the scaling in terms of $h$, simplifies to:

\begin{equation}
    \int_{\Omega} \omega (q) d r = \frac{1}{h^d}.
\end{equation}

The compact support ensures that the computational cost is reduced by limiting the number of neighboring particles that contribute to the interactions of a particle, as interactions are only computed within a finite range defined by $kh$. This is a key feature that makes the SPH method computationally feasible for large-scale problems such as tsunami and coastal wave simulations.

\subsubsection{Smoothing Kernel}
\vspace{0.15cm}
The performance of an SPH model depends heavily on the choice of the smoothing kernel. Kernels are expressed as a function of the non-dimensional distance between particles ($q$), given by $q=\frac{r}{h}$, where $r$ is the distance between any two given particles $a$ and $b$, and the parameter $h$ (the smoothing length) controls the size of the area around particle $a$ in which neighbouring particles are considered. 

Simply, the kernel function determines how much influence neighbouring particles have on the properties of a given particle. Generally, this influence is larger when the particles are coincident and the influence decreases smoothly as distance from the given particle increases, up to the smoothing length, after which particles have no influence.

Many smoothing kernels have been proposed and evaluated in the literature \citep{dehnen2012improving}, but in DualSPHysics, either a Cubic Spline kernel \citep{monaghan1985refined} or a Quintic kernel (known as a Wendland Kernel \citep{wendland1995piecewise}) are used widely. 

In this research, we use the Wendland kernel throughout due to its extensive application in coastal engineering to date. The Wendland kernel is defined as:

\begin{equation}
    W(r,h) = \alpha_D \left( 1 - \frac{q}{2} \right)^4 (2q + 1), \qquad 0 \leq q \leq 2
\end{equation}

\noindent where $\alpha_D = \frac{10}{7 \pi h^2}$ when simulating in 2D, or $\alpha_D = \frac{1}{\pi h^3}$ in 3-D. The factor $\alpha_D$ acts as a normalization constant to ensure mass conservation. In DualSPHysics, the smoothing length $h$ is calculated using a coefficient (\textit{coefh}) as an input parameter, determining the smoothing length as:

\begin{equation}
    h = \textit{coefh} \sqrt{3 dp^2}
\end{equation}

\noindent in 3-D simulations. According to the DualSPHysics guide, a typical value of $\textit{coefh} = 1$ is used, though for wave propagation simulations, a range of $1.2 \leq \textit{coefh} \leq 1.5$ is recommended. As a result, throughout this paper, we use a \textit{coefh} value of 1.5. We also tested a range of values of \textit{coefh}, including 1.2 and 1.5, to validate this choice for the specific application of tsunami runup and wave-structure interaction, which is further explained in this paper through the interpretation of results for Case II.

\subsubsection{Density and Pressure Calculation}
\vspace{0.15cm}
In SPH, the density at a particle, $\rho_i$, is computed as a weighted sum of the masses of neighbouring particles:

\begin{equation}
    \rho_i = \sum_j m_j W(r_{ij}, h)
\end{equation}

\noindent where $\sum_j$ is the sum of neighbouring particles ($j$), $m_j$ is the mass of neighbouring particles, and $W(r_{ij},h)$ is the smoothing kernel function based on the distance between particles $i$ and $j$.

Pressure is related to the density ($\rho_i$) using an equation of state (EOS), typically the Tait equation for weakly compressible fluids:

\begin{equation}
    P_i = \frac{c^{2}_s \rho_0}{\gamma} \left[ \left( \frac{\rho}{\rho_0} \right)^\gamma - 1 \right]
\end{equation}

\noindent where $P_i$ is pressure at particle $i$, $B$ is a constant that determines fluid compressibility, $\rho_0$ is the reference density, $\gamma$ is the polytropic index (usually 7 for water), and the numerical speed of sound is defined as:

\begin{equation}
    c_s = \sqrt{\partial P / \partial \rho}
\end{equation}

It is important to note that this numerical speed of sound is different to the physical speed of sound. This allows for larger time steps within the explicit time integration than would be possible with a physical speed of sound \citep{dominguez2022dualsphysics}.

\subsubsection{Momentum and Navier-Stokes Equations}
\vspace{0.15cm}
The momentum equation in SPH discretises the Navier-Stokes equations for fluid dynamics. The acceleration of a particle is given by the sum of forces acting on it, including pressure forces, viscous forces, and external forces (e.g., gravity):

\begin{equation}
    \frac{dv_i}{dt} = %
    - \sum_j m_j %
    \left( \frac{P_i}{\rho_{i}^2} + \frac{P_j}{\rho_{j}^2} \right) %
    \nabla W(r_{ij}, h) + \mu \sum_j m_j %
    \frac{v_i - v_j}{\rho_j} \nabla^2 W(r_{ij},h) + g
\end{equation}

\noindent where $v_i$ is the velocity of particle $i$, $P_i$ and $P_j$ are the pressures at particles $i$ and $j$, $\nabla W(r_{ij}, h)$ is the gradient of the smoothing kernel, $\mu$ is the dynamic viscosity, and $g$ is the external force per mass unit (in the case of this work, gravity). 

The continuity and momentum equations can be written in Lagrangian form as follows:

\begin{equation}
    \frac{d \rho}{d t} = -\rho \nabla \cdot v    
\end{equation}

\noindent and

\begin{equation}
    \frac{d v}{d t} = - \frac{1}{\rho} \nabla P + \Gamma + f
\end{equation}

\noindent respectively, where $d$ denotes the total or material derivative, v is the velocity vector; $\rho$ is density; $P$ is pressure; $\Gamma$ represents the dissipation terms; and $f$ represents acceleration due to external forces. 

\subsubsection{Time Integration}
\vspace{0.15cm}
DualSPHysics utilizes either a Verlet time integration scheme or a symplectic position Verlet time integrator scheme. All simulations within this work employ the original Verlet scheme. The governing equations are succinctly expressed as follows:

\begin{equation}
    \frac{dv_a}{dt} = F_a ; %
    \frac{d \rho_a}{dt} = R_a ; %
    \frac{d r_a}{dt} = v_a
\end{equation}

These equations are integrated over time using a Verlet-based scheme, which is commonly preferred due to its low computational cost and second-order accuracy in space, eliminating the need for multiple calculation steps within iteration intervals. The weakly compressible SPH variables are computed according to:

\vspace{-1cm}
\begin{gather}
    v_{a}^{n+1} = v_{a}^{n-1} + 2 \Delta t F_{a}^{n}; \\
    r_{a}^{n+1} = r_{a}^{n} + 2 \Delta t v_{a}^{n} + %
    \frac{1}{2} \Delta t^2 F_{a}^{n}; \\
    \rho_{a}^{n+1} = \rho_{a}^{n-1} + 2 \Delta t R_{a}^{n};
\end{gather}

DualSPHysics employs a variable time step, which depends on the Courant-Friedrichs-Lewy (CFL) condition, the forcing terms, and the viscous diffusion term. The minimum time step is calculated at each time step based on the velocity and force magnitudes per unit length.

\subsubsection{Boundary Conditions}
\vspace{0.15cm}
DualSPHysics describes boundaries as ``a set of particles that are considered as a separate to the fluid particles" \citep{DualSPHysics}, and several types of boundary conditions are implemented within the code. This is not an exhaustive explanation of boundary conditions, but cover those used within this work.

\textbf{Dynamic boundary conditions (DBC)} are fixed particles with density computed from the continuity equation and pressure from the equation of state and are the default boundary condition used in DualSPHysics. DBC boundary particles are fundamentally treated the same as fluid particles, other than that they do not move according to the forces exerted on them and remain fixed, or move according to imposed or specified motion (for example, the movement of a wavemaker paddle). 

As a particle moves toward a fixed DBC boundary, the boundary begins to fall within the smoothing kernel and the density of the particle concerned begins to increase, resulting in an increase in pressure. This pressure increase results in a repulsive force due to the pressure term in the momentum equation. 

Using a DBC boundary sometimes results in a non-physical interaction between the fluid and boundaries and this boundary condition can sometimes generate minor leaks of fluid particles through the solid particles, especially at sharp edges and corners, and can result in excessive dissipation. This boundary condition and its formulation can be found in \citet{cabrera2007boundary}.

\textbf{Modified dynamic boundary conditions (mDBC)} works in a similar method to DBC in that the boundary particles are arranged in the same way with the boundary interface being half of a particle spacing ($dp$) between the fluid and solid particle. In mDBC, during the case setup, for each boundary particle, a ghost node is projected into the fluid, mirrored across the boundary interface. The boundary particles store fluid properties using values calculated at the ghost node within the fluid domain. As fluid properties are stored and computed within boundary particles, the direction of the velocity is reversed and allows for the velocity of the fluid to be slowed the velocity of fluid particles as it approaches the boundary interface. 

Using mDBC has been demonstrated to reduce the gap between fluid particles and boundaries that has been shown to develop using DBC boundary conditions. It is also demonstrated to create a slip-free boundary interface by giving boundary particles the same tangential velocity found at its respective ghost node. This reduces dissipation and improves its ability to be used to simulate the width-wise span of a wavetank without excessively reducing wave propagation through energy dissipation. For this reason, and to be able to simulate flows around structures without a non-physical gap between the obstacles and the fluid particles, the mDBC boundary condition is used in this work. Further details of the formulation of mDBC and its validation can be found in \citet{english2019correction}.

\hspace{1.5cm}

\section{Case I: Idealised Tsunami Induced Currents in Port} \label{sec:CaseI_Breakwater}
Before using SPH to simulate runup and inundation, we validate its use for simulating turbulent fluid flows around coastal structures. In this case, we recreate the experimental case presented in \citet{borrero2015tsunami}, which represents a simplified pier commonly found in urbanised ports to protect public moorings and shipping berths from sea swell. 

This case setup is designed to validate the SPH method for simulating turbulent flows around structures, which is critical when modelling tsunami flows. Accurate validation will build confidence for extracting outputs from a real-world case inclusive of buildings and structures during tsunami inundation flows, and for models with increasing complexity.

\subsection{Case Setup and Configuration}

\begin{figure}[htbp]
\centering
\includegraphics[width=\linewidth]{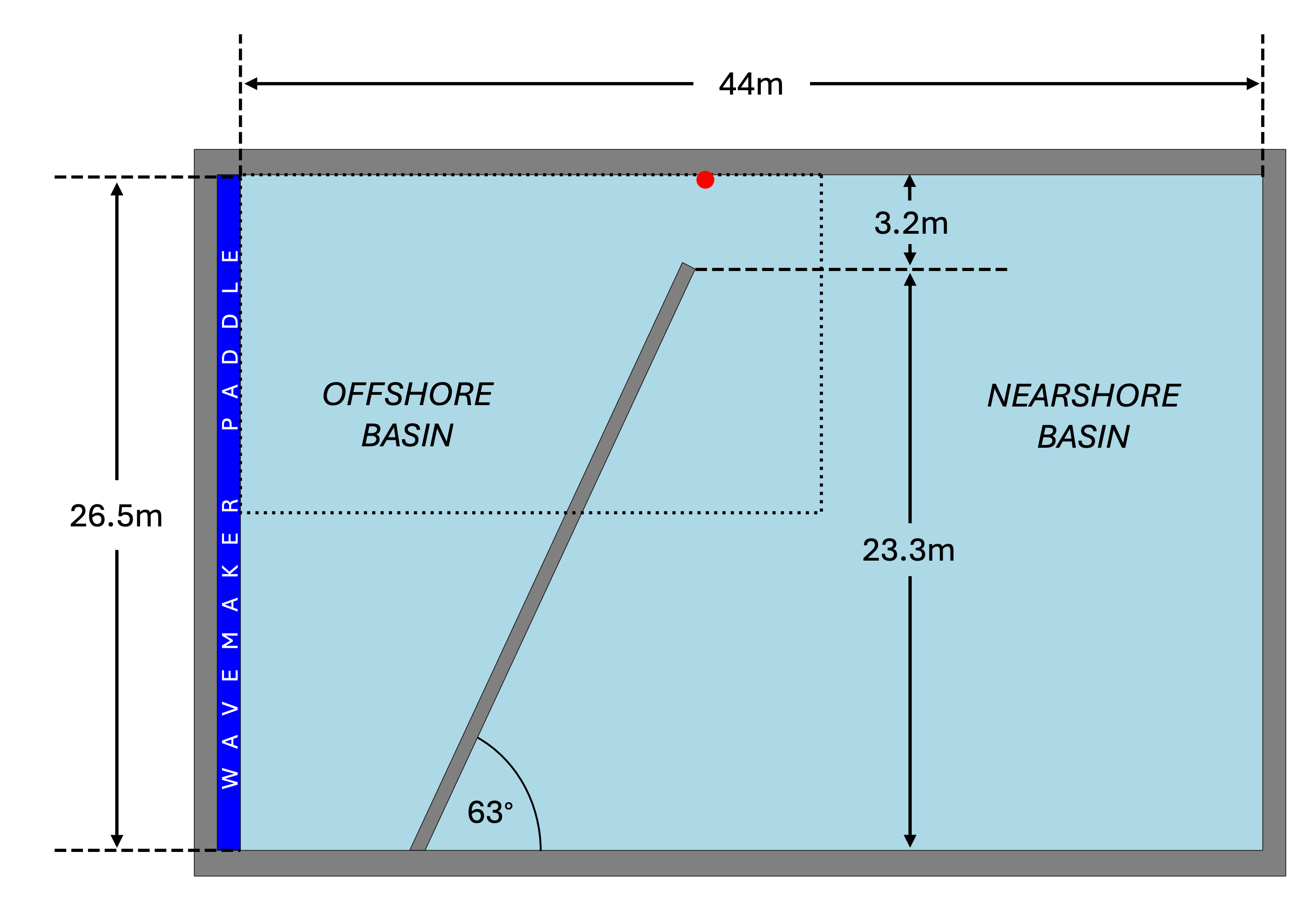}
\caption{Case setup for port breakwater experimental design outlined in \citet{borrero2015tsunami}. The wavemaker paddle is shown on right side of the tank in dark blue. Red dot on top edge represents virtual gauge location. Dotted area in top left part of the domain defines the subdomain shown in Figure \ref{fig:breakwatervelocityvectors}.}
\label{fig:breakwatersetup}
\end{figure}

The experimental setup outlined in \citet{borrero2015tsunami} is replicated in DualSPHysics with the same dimensions (Figure~\ref{fig:breakwatersetup}). The wave tank is 44 m x 26.5 m, with a paddle wavemaker on the left vertical edge. A protruding breakwater extends from the lower edge of the domain at an angle of 60$^\circ$, leaving a gap of ~3.2 m between the breakwater and the edge of the domain. The water level is set to 0.55 m depth. Therefore, this case represents a typical 15 m deep port at approximately 1:30 scale. A virtual wave gauge is placed in the gap between the breakwater and wall of the tank to extract a velocity timeseries at the same location as acoustic Doppler velocimeter (ADV) instruments used in the experimental setup.

\begin{figure}[htbp]
\centering
\includegraphics[width=0.65\linewidth]{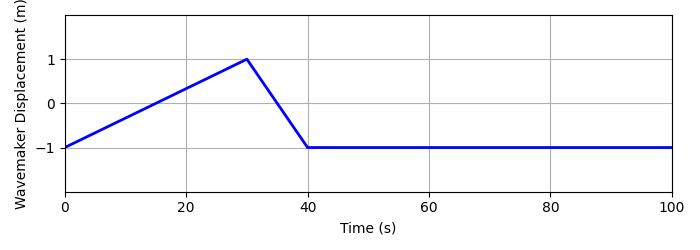}
\caption{Wavemaker paddle displacement for port breakwater experimental case outlined in \citet{borrero2015tsunami}, used as an input to the breakwater SPH simulation.}
\label{fig:breakwaterinput}
\end{figure}

A particle resolution ($dp$) of 0.05 m was chosen through a convergence test using values of $dp$ between 0.01 and 0.1 m. This test showed that for $dp < 0.05 m$ there was no significant increase in model convergence towards experimental data despite the increased computational cost. The total simulated time was 200 seconds. Default values were used for density, gravity, smoothing length coefficient, and viscosity treatment; key parametrisation for this case is given in Table~\ref{tab:breakwater_params}. The wavemaker velocity profile is shown in Figure~\ref{fig:breakwaterinput}.

\begin{table}[ht]
    \centering
    \caption{Key parameters used within DualSPHysics for the breakwater case}
    \begin{tabular}{lc} 
        \toprule
        Parameter & Value \\ 
        \midrule
        Number of Fixed Particles, $n_{fixed}$ & 1,433,209   \\
        Number of Moving Particles, $n_{moving}$ & 106,200   \\
        Number of Fluid Particles, $n_{fluid}$   & 4,675,960   \\
        Total Particles, $n_{total}$ & 6,215,369   \\
        Dimensions & Three-Dimensions \\
        Particle Resolution, $dp$ (m) & 0.05   \\
        Viscosity Treatment & Artifical Viscosity ($a$ = 0.01) \\
        Smoothing Kernel & \citet{wendland1995piecewise} \\
        Smoothing Length Coefficient & 2.598 \\
        Density (kg/m$^{3}$) & 1026 \\
        Gravity (m/s$^{2}$) & 9.81 \\
        CFL Number & 0.2 \\
        Simulated Time (s) & 200 \\
        \bottomrule
    \end{tabular}
    \label{tab:breakwater_params}
\end{table}

\subsection{Results}
\begin{figure}[htbp]
\centering
\includegraphics[width=0.75\linewidth]{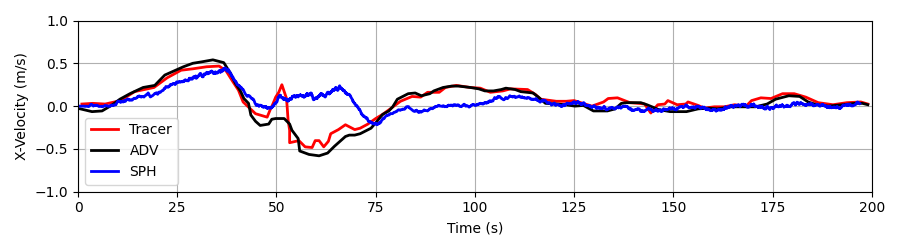}
\caption{X-component velocity gauge outputs from experimental and simulated results. The blue line shows the output from the virtual gauge within the SPH simulation. The red and black lines are experimentally-derived velocity outputs measured using the nearest tracer to the gauge location and acoustic Doppler velocimetry, respectively.}
\label{fig:breakwateroutput}
\end{figure}

The SPH solver, DualSPHysics (V5.4), was run on a single NVIDIA Tesla V100 (32GB) GPU, with a wall clock time of 14.14 hours. We observe the output from the velocity gauge in comparison to the experimental data in Figure~\ref{fig:breakwateroutput}. When comparing against the experimental results collected both from tracers on the free surface and ADV instrumentation, we can see reasonably good agreement, especially with the first wave arrival time and maximum amplitude. There is some separation between modelled and experimental results at the velocity gauge on the negative velocity phase as the wavemaker is quickly retracted.

\begin{figure}[htbp]
\centering
\includegraphics[width=0.5\linewidth]{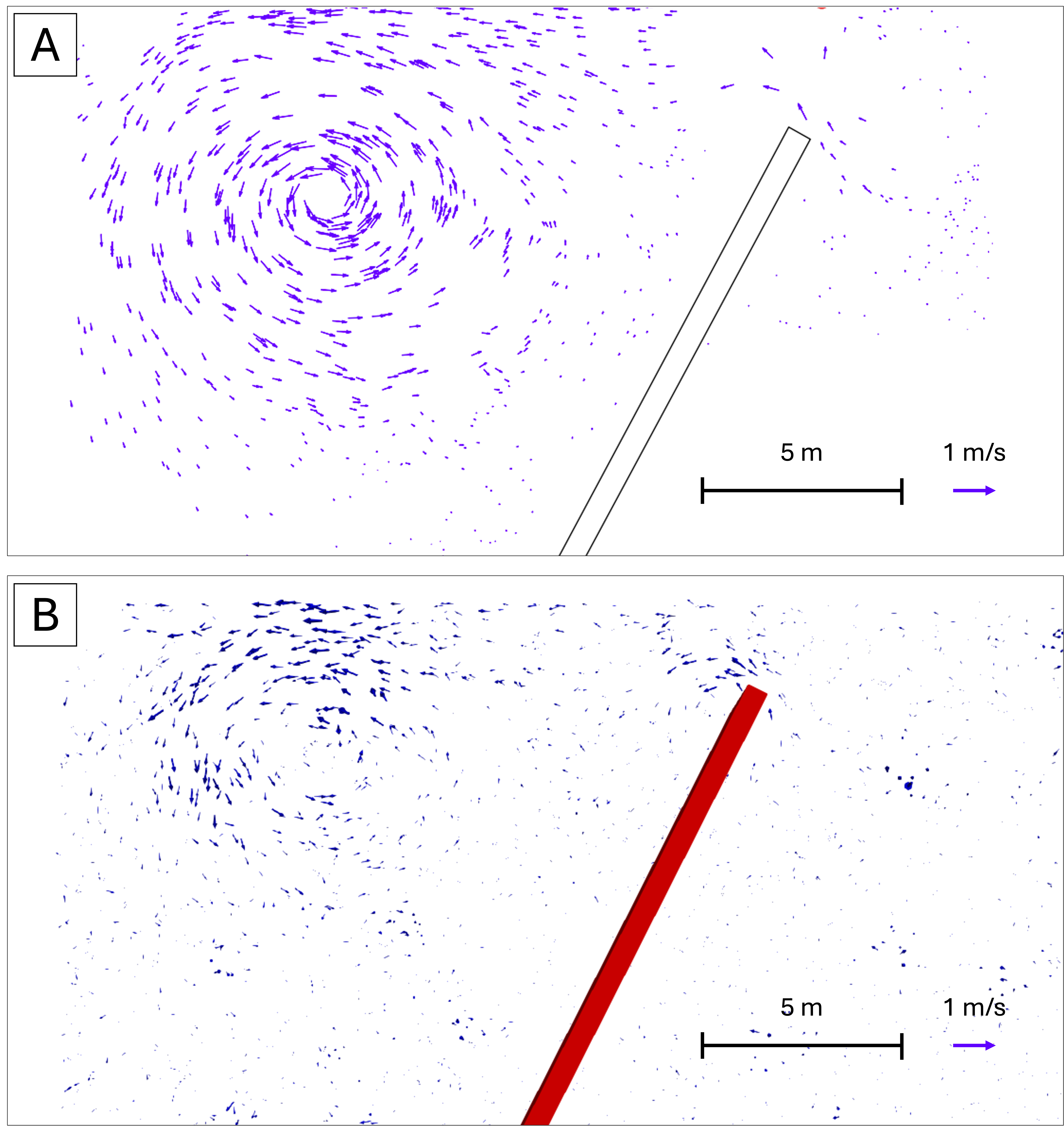}
\caption{Velocity vectors at $t=2.2$ minutes for (a) experimental setup \citep{borrero2015tsunami} and (b) SPH simulation. Figure extent and position relative to whole domain is shown in Figure~\ref{fig:breakwatersetup}.}
\label{fig:breakwatervelocityvectors}
\end{figure}

We can use the wider flow field to assess the extent to which the simulation is modelling the turbulent coherent structures resulting from flow around objects, such as the breakwater in this case. An advantage to the use of Lagrangian methods (such as SPH) is the ability to track the movement of single particles and visualise velocity vectors to then compare to experimental tracer data. In~Figure~\ref{fig:breakwatervelocityvectors}, we can visually compare the presence of an anti-clockwise rotating vortex in the same location. When comparing the magnitude of the velocity vectors between experimental and modelled outputs, we also find similar values. As is indicated in Figure~\ref{fig:breakwateroutput}, following the surge in fluid flow, the drawback and seaward flow is captured to a lesser magnitude in SPH, and so this is likely the reason for the marginally smaller velocity vectors also seen in the SPH results on the seaward side of the breakwater. We further discuss this phenomenon in our discussion.

\section{Case II: Seaside, Oregon} \label{sec:CaseII_Seaside}

To progress the model development closer to a final real-world case, we create a numerical replica of the experimental study of tsunami flows across Seaside, Oregon, as published in \citet{park2013tsunami}. As well as presenting experimental results, \citet{park2013tsunami} also validates the COULWAVE wave model against this data, making it an interesting case both to validate SPH but also compare against other numerical models.

\begin{figure}[htbp]
\centering
\includegraphics[width=0.9\linewidth]{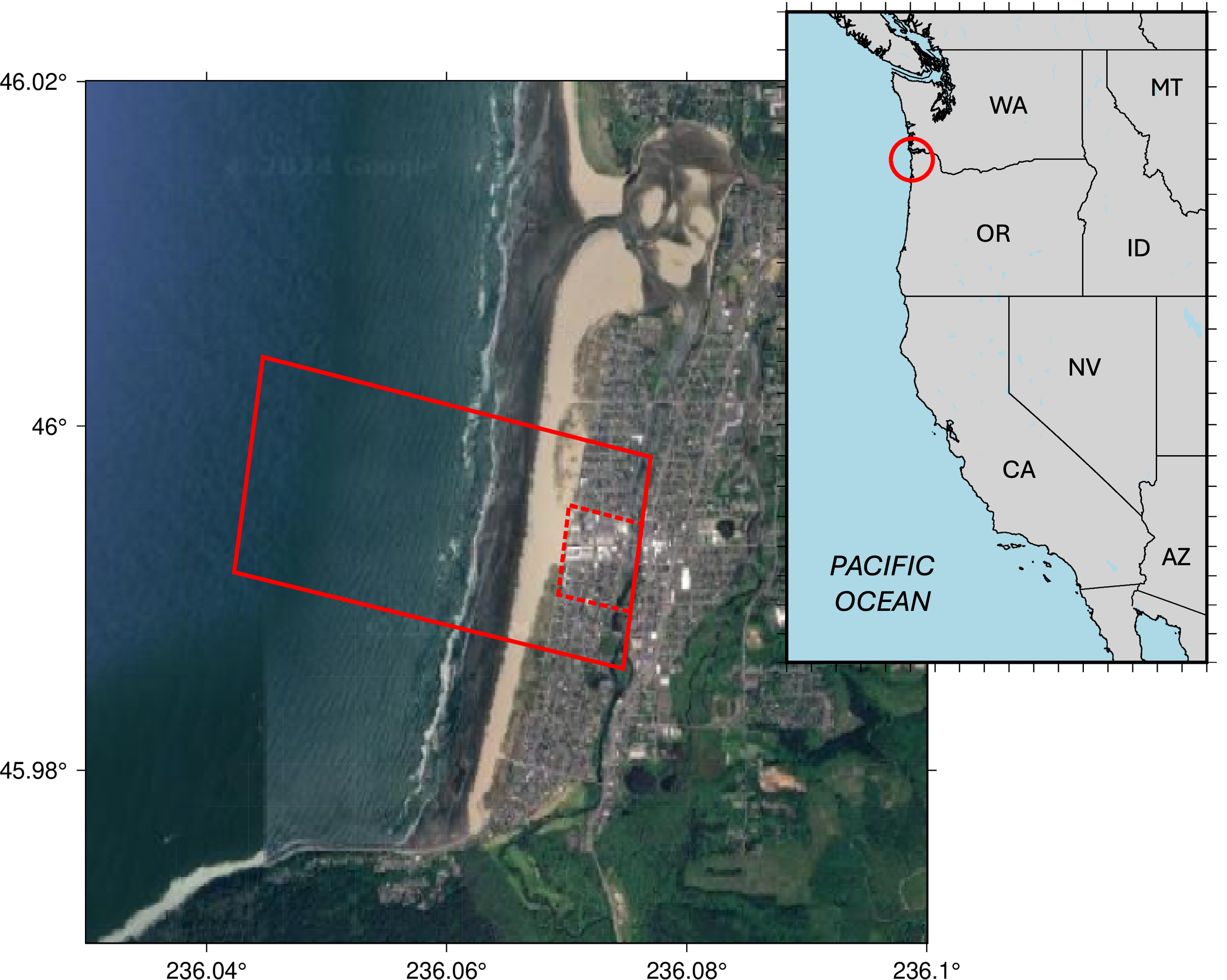}
\caption{Location of Seaside, Oregon case. Red extent in main figure shows the extent of the SPH and experimental modelling domain, and inner dotted extent denotes the extent of modelled buildings and structures. Inset shows location of Seaside relative to the western coast of the United States of America. Basemap imagery is sourced from Google Maps Imagery/Landsat.}
\label{fig:seasidelocation}
\end{figure}

The city of Seaside is selected as a site due to its proximity to the Cascadia Subduction Zone (CSZ) which runs along the western coast of the United States of America and Canada, making this coastline prone to tsunami events. It is also a useful case for observing tsunami flows around structures due to the range of commercial and residential buildings close to the waterfront \citep{park2013tsunami}. The location of and extent of this case are shown in Figure~\ref{fig:seasidelocation}.

\subsection{Case Setup and Configuration}

\begin{figure}[htbp]
\centering
\includegraphics[width=0.9\linewidth]{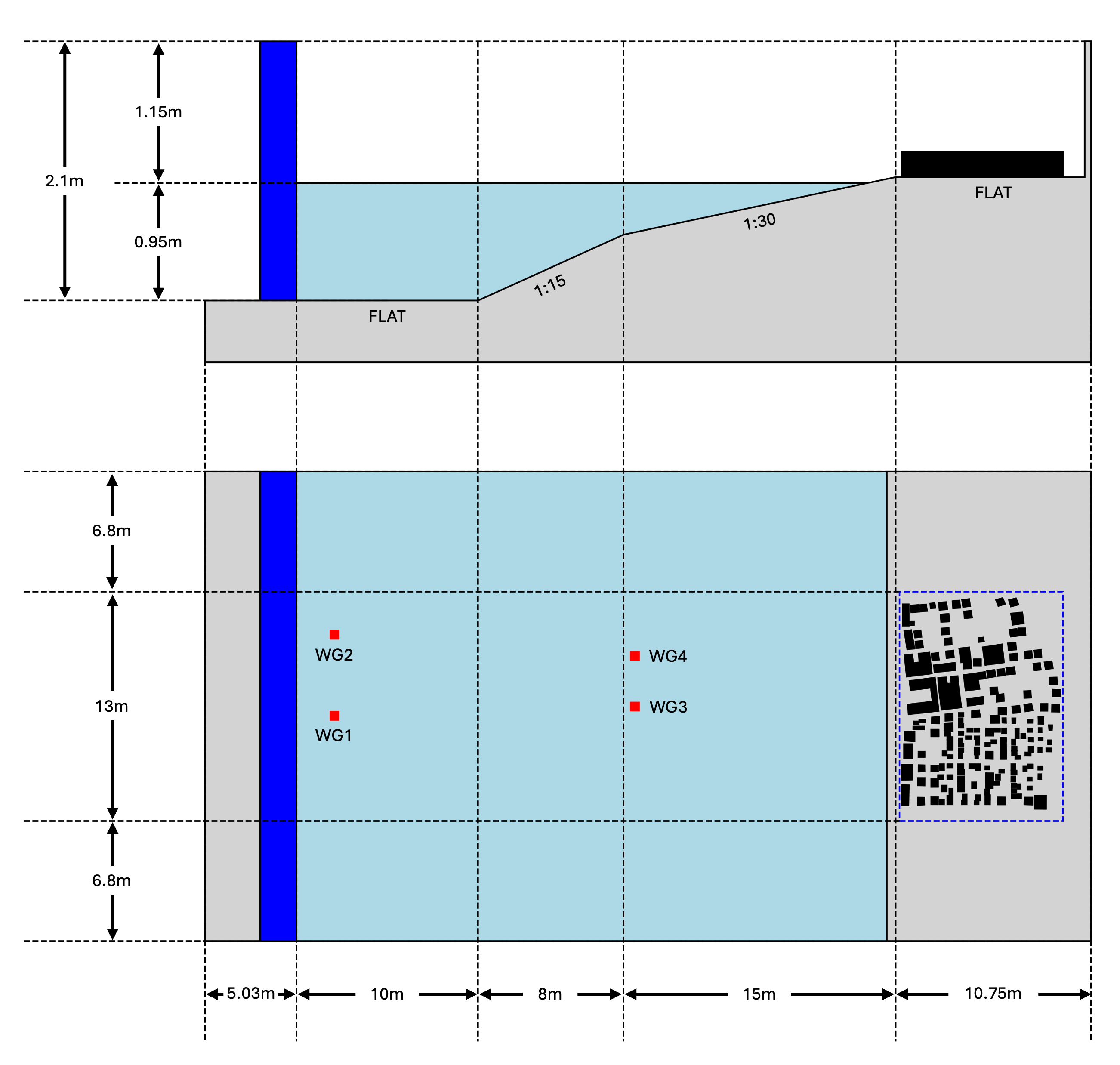}
\caption{Case setup for Seaside, Oregon scale model. Light blue areas represent fluid particles and grey areas represent the solid boundary particles of the wave basin. Dark blue polygon represents the piston wavemaker. Black polygons on the nearshore are scale models of real-world buildings in Seaside. Red squares in the plan view represent virtual and experimental wave gauges.} 
\label{fig:seasidedimensions}
\end{figure}

The Seaside scale wave tank and numerical setup is 48.8 m in length, 26.5 m in width, and 2.1 m deep. The water depth is set at 0.95 m deep. The tank has a piston wavemaker at the seaward edge, and idealised bathymetry consisting of three sections of varied slope angle. The topography in this case is completely flat, with an array of buildings modelled in the centreline of the modelling domain, with space left at both sides of the buildings along the longshore axis to mitigate influences caused by interactions with the tank wall. A graphical visualisation of the model setup and gauges can be found in Figure  \ref{fig:seasidedimensions}.

\begin{figure}[htbp]
\centering
\includegraphics[width=0.8\linewidth]{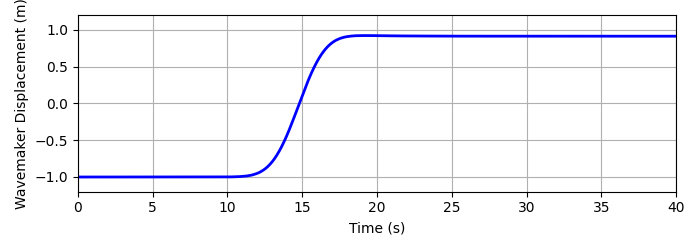}
\caption{Wavemaker piston paddle displacement used in the experimental case setup and as an input to the SPH simulation.}
\label{fig:seaside_paddleinputs}
\end{figure}

We use the same wavemaker piston movement as is used in \citet{park2013tsunami} as an input to the SPH setup (Figure~\ref{fig:seaside_paddleinputs}). The initial condition of the wavemaker places the leading edge of the paddle at $x=0m$ and moving into the domain over a distance of close to 2 m, and so the displacement values of this input timeseries were normalised to start from a displacement of 0 m. This avoids a sudden drawback of the wavemaker at the first timestep to meet the required first displacement value of -1 m, this phenomenon is discussed later in this work.

We conduct a number of sensitivity tests for the input parameters that, given the nature of the case, would be particularly influential on the propagation, runup, and inundation of the tsunami wave in this case setup. Using the experimental data as a truth, we identify appropriate values for these key simulation parameters to then use when setting up future cases where experimental or historical data is not available. We aim to calibrate the following parameters through this process: $dp$ (particle resolution), CFL number, smoothing length coefficient, and boundary conditions. These tests result in the choice of parameters for the final case setup defined in Table~\ref{tab:seaside_params}. We only explored scenarios and offer a simple tuning as the sensitivities (see below) are either negligible or clearly converging in one direction. A more advanced Bayesian calibration (see e.g. \citet{guillas2014bayesian} for a CFD calibration) could be carried out to formally explore a wider range of parameters for finer tuning. However this task is beyond the scope of this first paper demonstrating agreement against observations under a reasonable set of choices for the parameters.

\begin{table}[ht]
    \centering
    \caption{Parameters used as inputs to DualSPHysics for the Seaside, Oregon case}
    \begin{tabular}{p{7cm}p{5cm}} 
        \toprule
        Parameter & Value \\ 
        \midrule
        Number of Fixed Particles, $n_{fixed}$ & 4,378,178 \\
        Number of Moving Particles, $n_{moving}$ & 223,860 \\
        Number of Fluid Particles, $n_{fluid}$   & 3,905,824 \\
        Total Particles, $n_{total}$ & 8,507,862 \\
        Dimensions & Three-Dimensions \\
        Particle Resolution, $dp$ (m) & 0.05   \\
        Viscosity Treatment & Laminar Viscosity with Sub-Particle Scale Turbulence \\
        Kinematic Viscosity (m\textsuperscript{2}s) & $10^{-6}$ \\
        Smoothing Kernel & \citet{wendland1995piecewise} \\
        Smoothing Length Coefficient & 2.598 \\
        Density (kg/m$^{3}$) & 1026 \\
        Gravity (m/s$^{2}$) & 9.81 \\
        CFL Number & 0.2 \\
        Simulated Time (s) & 40 \\
        \bottomrule
    \end{tabular}
    \label{tab:seaside_params}
\end{table}

\subsection{Results}
\begin{figure}[htbp]
\centering
\includegraphics[width=0.65\linewidth]{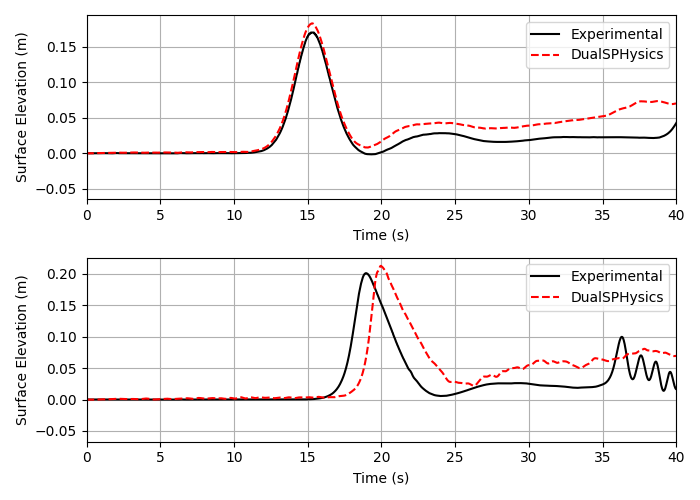}
\caption{Wave gauge output comparison between DualSPHysics output and experimental results from \citet{park2013tsunami}, for (top) WG1 and (bottom) WG3.}
\label{fig:seaside_WGvalidation}
\end{figure}

This case is run on a single NVIDIA Tesla V100 (16GB) GPU, with a wall clock time of 74.6 hours. When we compare the offshore wave gauge outputs produced by the SPH simulations to that of the experimental data, we see a good agreement both in phase difference and amplitude (Figure~\ref{fig:seaside_WGvalidation}). The comparison is especially good for the wave gauge closest to the wave paddle, and as the wave slows and shoaling occurs, the phase difference between the two cases increases. However, despite that the maximum phase difference between the experimental and simulated observations, we observe an accuracy in peak amplitude of within 2 cm of the experimental data. 

\begin{figure}[htbp]
\centering
\includegraphics[width=\linewidth]{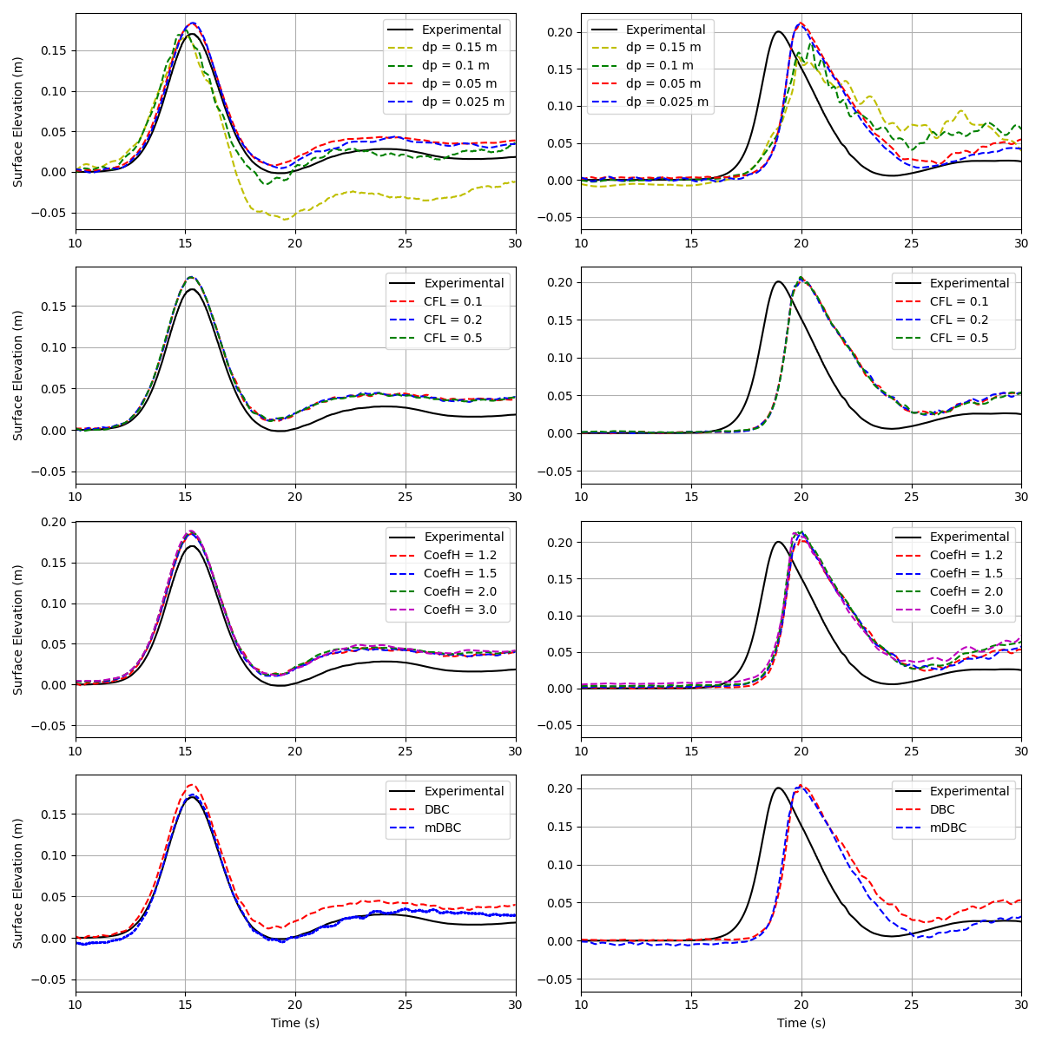}
\caption{Simulation parameter sensitivity analysis for Seaside, Oregon DualSPHysics case, and comparison to experimental results from \citet{park2013tsunami} for (a-b) particle resolution, $dp$, (c-d) CFL number, (e-f) smoothing length coefficient, $\text{CoefH}$, and (g-h) boundary condition. Subplots (a), (c), (e), and (g), show outputs from wave gauge 1 (WG1) and subplots (b), (d), (f), and (h) show outputs from wave gauge 2 (WG2). Note that in subplots (a-b), the surface elevation is normalised between simulations of varied resolution, due to the initial water height varying slightly due to variance where the resolution determined the number of discrete particles that can be placed within the water column height.}
\label{fig:seaside_calibration}
\end{figure}

Through the process of calibrating the key input parameters of $dp$, boundary condition, smoothing length, and CFL number (Table~\ref{tab:calibration_params}) in 2D cross sections of the Seaside case, we find that the most sensitive parameter is particle resolution. We find that particle resolution must be sufficiently high to capture the breaking wave and turbulent flows, and increasing particle resolution results in a smoother output that is closer to the experimental output values (Figure~\ref{fig:seaside_calibration}). We note that the alteration of the CFL number had negligible impact, but it is perhaps worth mentioning that this is likely due to the fact the wave gauges were in deeper water and was not capturing especially complex and rapid flows. Similar is true for smoothing length coefficient, until we begin observing the reflections from the beach towards the end of the timeseries outputs; this is to be expected on the basis that particle energy dissipation is heavily reliant on the shape and size of the smoothing kernel that lies at the core of the SPH numerical scheme, and so changing this would likely have a significant influence over energy dissipation \citep{colagrossi2013smoothed}. While mDBC shows a marginal improvement over the DBC, this comes at the cost of having to define surface normals for the geometry, which is challenging for more complex geometries. As is well documented in SPH research, a higher particle resolution generally leads to improved agreement with experimental data, however we must manage this within computational limits. We observe that continuing to increase particle, at a point, will begin to show little significant improvement in result, whilst dramatically increasing simulation wall clock time. On this basis, we aim to use the largest resolution possible to both capture the wave dynamics and outputs of interest, but also maintain the highest levels of computational efficiency possible.

\begin{table}[h]
    \centering
    \caption{Table of parameters calibrated using the Seaside, Oregon Case setup}
    \begin{tabular}{p{2.5cm}p{10cm}} 
        \toprule
        Nomenclature & Parameter \\ 
        \midrule
        $dp$ & Defines the distance between particles \\
        $CFL$ & CFL number coefficient to multiple $dt$ (time step) \\
        $CoefH$ & Coefficient to calculate the smoothing length, $h = \textit{CoefH} \cdot \sqrt{3 \cdot dp^2}$ in 3D \\
        DBC/mDBC & Boundary condition, Dynamic Boundary Condition (DBC) or Modified Dynamic Boundary Condition (mDBC) \\
        \bottomrule
    \end{tabular}
    \label{tab:calibration_params}
\end{table}

\section{Case III: South Cilacap, Central Java} \label{sec:CaseIII_Cilacap}
\begin{figure}[htbp]
\centering
\includegraphics[width=\linewidth]{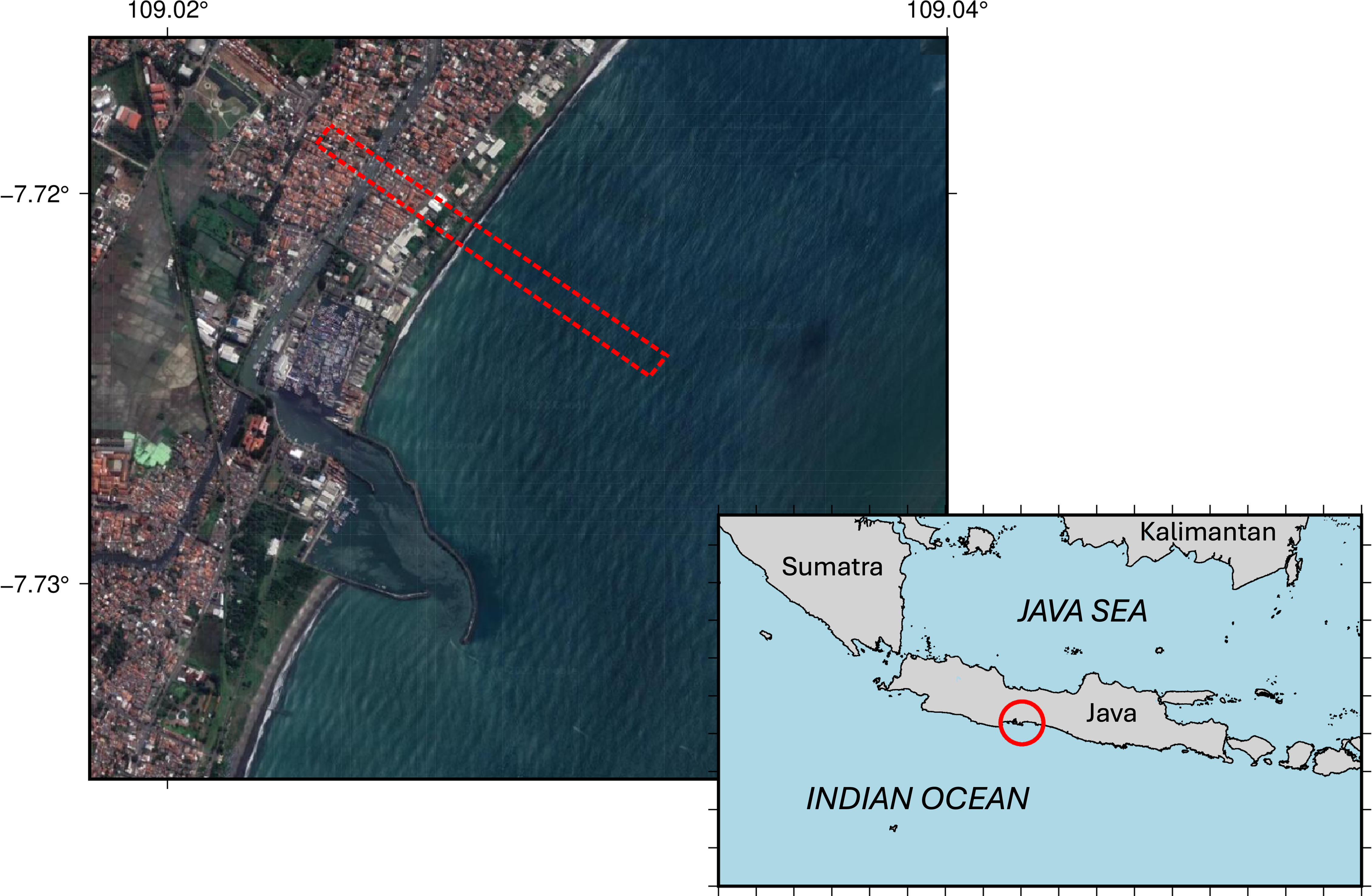}
\caption{Location of Cilacap model setup. Red dashed domain in main figure shows the SPH modelling domain. Inset shows the location of the main figure (red circle) relative to Java, Sumarta and Kalimantan Islands.}
\label{fig:cilacaplocation}
\end{figure}

Finally, we combine real-world bathymetry and topography with the use of coastal urban structures. The case focuses on the city of Cilacap, located on the southern coast of Central Java, Indonesia (Figure~\ref{fig:cilacaplocation}). This location was chosen due to its exemplarity for tsunami risk assessments, as shown in \citet{salmanidou2021impact}. In order to simulate this case at a resolution that would capture turbulent processes around buildings and structures to an extent whereby the tsunami flow could progress between the buildings and provide meaningful outputs throughout the domain, we simulate a 150 m cross-section of the Cilacap coastline. 

This case builds on the validation and calibration of the use of SPH for simulating tsunami flows across urban topographies outlined in Sections \ref{sec:CaseI_Breakwater} and \ref{sec:CaseII_Seaside}, by serving as a test case for a scenario that does not have experimental or historical validation data. This is important to achieve as the benefit of this type of modelling for tsunami flows lies in its ability to test scenarios that have not yet occurred, and to then be developed to run ensembles of cases that can form probabilistic assessments of tsunami impact. 

\subsection{Case Setup and Configuration} \label{subsec:inletoutlet}

As this case is setup to represent an urban coastal region as closely as possible, it combines a range of real-world datasets. For the base of the tank, we use digitised charts containing bathymetric measurements from a collection of surveys that cross the study area \citep{navionics}. We also collect land topography data from \citet{srtm_usgs} and use this for the terrestrial wavetank base by aligning and merging it with the edge of the bathymetric dataset. To reconstruct the buildings, we take the dataset from OpenStreetMap where building shapes and heights can be extracted \citep{OpenStreetMap}. Where this data was not available, the outlines of buildings were extracted manually from optical satellite imagery, and their heights approximated through comparison to known structures using street-level imagery. In practice, as buildings in this area were generally of similar heights, this did not result in a large range of building heights. 

We then extract a cross section to include the shallow bathymetry, topography, an array of buildings scaled and oriented in their real-world layout, and a river trench towards the inland part of the domain. The seaward end of the modelling domain has a depth of 11 m, and the depth of the river trench towards the landward edge of the domain was 3.5 in depth, with sloping sides. There is an inlet/outlet boundary condition created at the seaward end of the domain with a flat base created by slightly smoothing the bathymetry at this deepest edge. 

\begin{figure}[htbp]
\centering
\includegraphics[width=\linewidth]{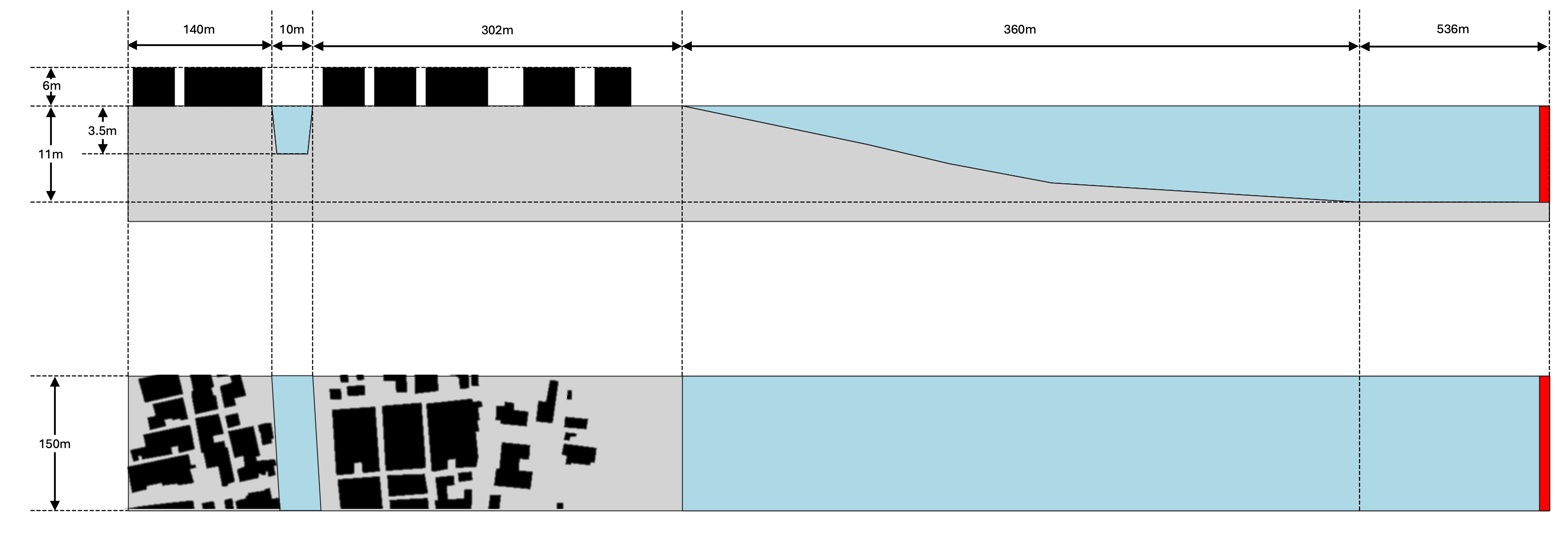}
\caption{Dimensions of the SPH domain recreating a section of Cilacap, Indonesia in side view. Black filled polygons represent buildings, blue represents fluid particles, and the red boundaries represent the inlet/outlet boundary domain. Note that bathymetry and topography is simplified for this diagram.}
\label{fig:cilacapdomain}
\end{figure}

The inlet/outlet boundary uses the numerical scheme outlined in \citet{verbrugghe2019nonlinear} to allow fluid particles to both enter and leave the simulation domain. This allows for long-wavelength waves to be simulated without the requirement of a very long wavetank or particles accumulating in the domain leading to nonphysical results. This inlet/outlet is a type of open boundary \citep[defined in ][]{tafuni2018versatile} that is implemented by a buffer zone of ghost particles upon which physical quantities such as velocity or surface elevation can be imposed. If these particles then cross the boundary edge of the domain, they are added, and if they leave the simulation domain into the buffer zone, they are removed. To avoid overconstraining the boundary buffer particles, resulting in non-physical results or instabilities, we provide the boundary with a velocity timeseries and allow it to compute the valid surface elevation. Preliminary investigations also studied the reverse, where only the surface elevation is given as an input but this gave poor results in the form of waves that propagated far too rapidly through the domain resulting in a wave height far higher than was realistic. 

For the purposes of testing this case, a synthetic tsunami event was generated with the seismic parameters given in Table~\ref{tab:cilacap_case_seismicparams}. Whilst the latitude and longitude were deliberately sampled as a location that would generate a wave likely to produce inundation in coastal Cilacap, the other parameters were calculated using scaling relationships and knowledge of regional tectonics. Fault width and length were calculated using the scaling relationships presented in \citet{wells1994new}. Using the latitude and longitude, we used the Slab2 dataset to compute dip, strike, and depth of the earthquake. Slab2 is a three-dimensional model that provides detailed information about the geometry of subducting tectonic plates, produced and maintained by the US Geological Survey \citep{slab2}.

\begin{table}[ht]
    \centering
    \caption{Seismic parameters for Cilacap test case.}
    \begin{tabular}{lc} 
        \toprule
        Parameter & Value \\ 
        \midrule
        Latitude, $y$ & -9.785 \\
        Longitude, $x$ & 110 \\
        Fault Length, $l$   & 75.813 km \\
        Fault Width, $w$ & 26.451 km \\
        Dip, $dip$ & 5.45$^{\circ}$   \\
        Rake, $rake$ & 90$^{\circ}$   \\
        Strike, $str$ & 277.09$^{\circ}$   \\
        Depth, $dep$ & 10.023 km   \\
        Moment Magnitude, $Mm$ & 7.26   \\
        \bottomrule
    \end{tabular}
    \label{tab:cilacap_case_seismicparams}
\end{table}

\begin{figure}[htbp]
\centering
\includegraphics[width=0.7\linewidth]{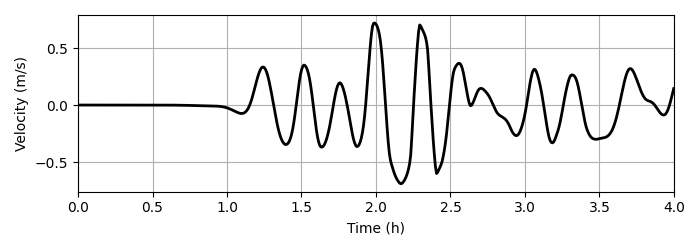}
\caption{DualSPHysics inlet/outlet boundary condition input velocity time series for Cilacap test case.}
\label{fig:cilacap_inputs}
\end{figure}

Using the MOST solver through the ComMIT model interface (v1.8.3), these input parameters were propagated across a set of three nested bathymetric grids (see Table~\ref{tab:cilacap_most_grids} for grid parameters) towards the area of Cilacap to generate a gauge output at the same location as the inlet/outlet boundary in the SPH domain. The MOST simulation across three nested grids was run through the ComMIT user interface on an Apple M1 chip, with a wall clock time of 55 seconds. A coarser resolution setup was used to focus on improving computational efficiency as capturing shallow water processes was not necessary, as only an offshore wave gauge was required to serve as an input to the SPH inlet/outlet boundary. An absorptive boundary was used along the 1 m contour of the MOST generation and propagation setup to avoid wasting computational effort computing runup and inundation. The resulting velocity input time series is shown in Figure~\ref{fig:cilacap_inputs}.

\begin{table}[ht]
    \centering
    \caption{MOST input grid parameters for Cilacap test case.}
    \begin{tabular}{cccccc}
        \toprule
        Grid ID & $X_{min}$ & $X_{max}$ & $X_{min}$ & $X_{max}$ & Resolution \\ 
        \midrule
        A & 101.9672 & 120.3422 & -14.0898 & -3.7898 & 2500 m \\
        B & 108.5992 & 109.4742 & -7.9500 & -7.5600 & 500 m\\
        C & 108.9311 & 109.1078 & -7.8063 & -7.6796 & 91 m\\
        \bottomrule
    \end{tabular}
    \label{tab:cilacap_most_grids}
\end{table}

For the setup of the simulation parameters, we use the parameters values calibrated in Section~\ref{sec:CaseII_Seaside}, noting that the case is created true to scale compared 1:50 scale used in the Seaside case. Using a true-to-scale setup removes the scale effects caused by altering the size of the domain and fluid flow obstacles (in this case, buildings). On this basis, the required particle resolution can be scaled also. The size of this case means calibrated $dp = 0.05 m$ is scaled up to $dp = 2.5 m$. However, when running the case at this resolution, it was insufficient to capture the waveform entering the domain from the inlet/outlet boundary. Being aware of the fact that the simulation time increases approximately to the power of four (once for each spatial dimension and once for the time step) as we decrease the value of $dp$, we conducted some calibration to reduce the value of $dp$ as far a possible whilst balancing computational demand and runtime. We consider a reasonable run time to be less than 24 hours. 

\begin{table}[ht]
    \centering
    \caption{Parameters used as inputs to DualSPHysics for the Cilacap case}
    \begin{tabular}{p{7cm}p{5cm}} 
        \toprule
        Parameter & Value \\ 
        \midrule
        Number of Fixed Particles, $n_{fixed}$ & 763,094 \\
        Number of Moving Particles, $n_{moving}$ & 0 \\
        Number of Fluid Particles, $n_{fluid}$   & 677,764 \\
        Total Particles, $n_{total}$ & 1,440,858 \\
        Dimensions & Three-Dimensions \\
        Particle Resolution, $dp$ (m) & 0.5 \\
        Viscosity Treatment & Laminar Viscosity with Sub-Particle Scale Turbulence \\
        Kinematic Viscosity (m\textsuperscript{2}s) & $10^{-6}$ \\
        Smoothing Kernel & \citet{wendland1995piecewise} \\
        Smoothing Length Coefficient & 1.5 \\
        Density (kg/m$^{3}$) & 1026 \\
        Gravity (m/s$^{2}$) & 9.81 \\
        CFL Number & 0.2 \\
        Simulated Time & 14400 s / 4 h \\
        \bottomrule
    \end{tabular}
    \label{tab:cilacap_params}
\end{table}

In order to view the outputs of this case, we construct a set of surface elevation gauges across the initially dry part of the domain (Figure~\ref{fig:cilacapdomain}). These wave gauges (WG1-WG4) allow us to view the surface elevation during the final stages of runup and the start of the inundation phase of the tsunami wave. The two buildings within this setup are also configured to act as force gauges, to allow us to quantify the force applied to buildings on impact of the wave. We additionally output the surface elevation at the inlet interface with the fluid to act as an 'input elevation' time series as this is actually computed based on the input velocity profile. 

\subsection{Results}
This case is run on a single NVIDIA Tesla A100 (40GB) GPU, with a wall clock time of 73.2 hours. It is also important to revisit the variable time stepping mentioned in Section~\ref{sec:sphformulation} that is used in DualSPHysics, and that run time will also increase in cases where larger velocity and force magnitudes exist, and vice versa.

\begin{figure}[htbp]
\centering
\includegraphics[width=\linewidth]{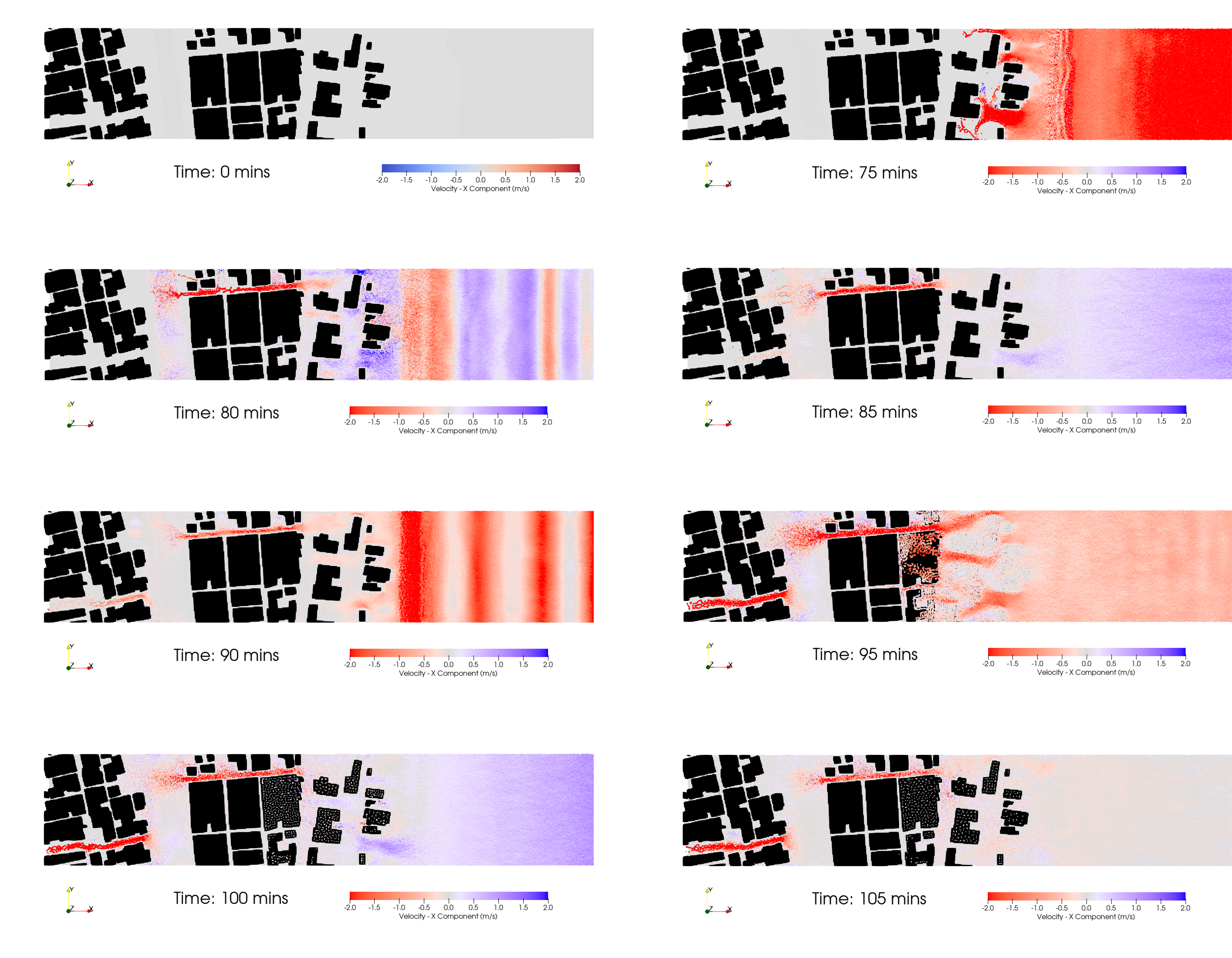}
\caption{Snapshot outputs of South Cilacap 3-D SPH case every 5 minutes for the initial wave impact and withdrawal. Colour scale shows particle velocity in the x-component. Bathymetry particles are shown in grey and buildings are shown in black.}
\label{fig:2DOutputSnapshots}
\end{figure}

This runtime also included the output of snapshots of particle location, density, and velocity magnitudes every 10 seconds of simulated time. A selection of these outputs can be seen in Figure~\ref{fig:2DOutputSnapshots}. Here we observe the gradually surging wave resulting from the long wavelength tsunami wave input. Following some minor oscillations in the x-component velocity, the arrival of the first peak then overtops the first building and drains into the cavity further inland where the river channel is flooded and its banks are breached. Finally, this fluid begins to leave the domain behind the second building. This landward building is deliberately left open to best simulate the fluid continuing to flow inland.

The tsunami wave takes approximately 3 mins from entering the seaward edge of the domain to crossing the initial condition shoreline. We can see that across the snapshot outputs shown in Figure~\ref{fig:2DOutputSnapshots}, there is an almost constant onshore flow despite the tsunami retreating at the coast. Although difficult to observe visually, this onshore flow does vary in velocity between 5 m/s and 0.5 m/s. This continuous onshore flow, just varying in magnitude, is likely influenced by the bathymetry and buildings creating a barrier for a strong offshore, and the fact that the inertia-driven flow onshore continues to progress unless slowed by external forces. The reversal of the source flow should not be misunderstood as exerting a negative pressure or suction on the inland fluid domain. Instead, it marks the cessation of the forcing from the sea, thereby removing the primary driver of inland advection without actively inducing a return flow.

When observing the particle velocity along the cross-shore component, we can see a major channelling effect of the tsunami wave propagating onshore between buildings. This flooding of the domain results in the inshore river breaching and creating a secondary channel along the path of least resistance, in this case, the widest available road. It is a reasonable assumption that this effect is exacerbated by the lack of flow through the narrower streets, prevented by the particle resolution of this case.

\section{Discussion}
This paper seeks to explore the SPH method and its applicability for simulating tsunami problems, developing in complexity using experimental test cases and validating against their results. Many existing tsunami simulators tend to exclude the presence of structures and obstacles and instead focus on bare earth topography/bathymetry, relying on friction coefficients as resistance to inundating fluid flow in place of land-based structures \citep{choi2012tsunami}. Using a Lagrangian particle-based method such as SPH allows for the potential to introduce obstacles and simulate flows with violent free-surface dynamics that otherwise be extremely challenging for traditional Eulerian mesh-based methods without requiring additional techniques.

The results of each of the two experimentally validated cases demonstrated well that the SPH method was easily capable of providing outputs similar to its experimental counterpart. Some variances did appear between the experimental and simulated results, as shown in the location of the eddy that formed on the offshore side of the breakwater demonstrating a more compact vortex core radius. However, the differences are arguably marginal and the location of these eddies was virtually identical across the experimental and simulated outputs. This eddy formation advancement seen in the experimental data when compared against the SPH simulated outputs is potentially due to the data acquisition latency involved in the capture and processing of tracer data or the accuracy of the experimental recording of time. However, this is not relevant in simulated time as an exact snapshot in time is far easier to achieve. When we observe the timeseries outputs, we do see a generally very positive agreement with the experimental data both in terms of first wave arrival and amplitude. It is likely that the deviation in the trough of this initial wave observed at the gauge output is due to the difference in wave energy absorbtion between the wavemaker in the experimental setting and the solid boundary particles used in the SPH setup.

As we progress onto Case II, waves are simulated to impact an array of simplified coastal structures after runup across simplified bathymetry. When comparing the wave gauges between experimental and simulated outputs (Figure~\ref{fig:seaside_WGvalidation}), there is an excellent agreement between the wave amplitude and phase close to the wavemaker. This allows us to have confidence that the SPH setup is effectively recreating the wave as it enters the domain, and deviation between the outputs further into the domain is as a result of the parameters and numerical scheme within the SPH setup, and not methodological uncertainties or systematic errors. As the wave progresses into the nearshore region at WG3, the wave profile is beginning to move out of phase, with the simulated wave propagating more slowly than the experimental case. This could be for a number of reasons; most likely resolution, or boundary conditions.

\subsection*{Particle Resolution, $dp$}
Particle resolution or particle spacing ($dp$) is a parameter that is extensivley discussed, calibrated, and validated across particle-based methods but in SPH specifically. It is widely known that using a smaller particle spacing, and thus generating a higher resolution case, is better at resolving complex and rapidly changing flows \citep{wang2024numerical}. However, this comes at the expense of computational efficiency. \citet{dominguez2021dualsphysics} demonstrates that regardless of the hardware used to run DualSPHysics, increasing the number of particles (which happens as a result of reducing particle spacing, $dp$) has a dramatic scaling effect on runtime. For example, running the DualSPHysics 2D dam break example case of 1 million particles using a Tesla V100 GPU takes approximately 2 hours, whereas running with 2 million particles takes 16 hours. It is therefore important to use the minimum resolution required to fully resolve the phenomena being simulated. 

\citet{roselli2018ensuring} and \citet{altomare2017long}, amongst others, suggest that particle spacing is set to allow for at least 5 particles to capture the largest wave height. Often written as $\frac{H}{dp} = 5$, where $H$ is the wave height, this is considered the suggested requirement to capture waves in SPH. That said, our results do demonstrate that the wave in Case II can be resolved well at $dp = 0.05$, which equates to roughly $\frac{H}{dp} = 3.4$. Also, when we half the particle spacing to $dp = 0.025$ in the calibration tests, which is $\frac{H}{dp} = 7.2$, shows no significant impact and does not change the phase difference between simulated and observed values. Halving $dp$ also causes the simulation runtime to increase from 96 hours to 768 hours (32 days). We therefore surmise that using $dp = 0.05$ and therefore $\frac{H}{dp} = 3.4$ is in fact sufficient in this case setup, and that it is possible to push the recommendations to ensure a minimum of $\frac{H}{dp} = 5$ in favour of computational efficiency. On this basis, it is also unlikely that particle resolution is a significant factor in the excessive dissipation of the wave as it propagates through the Seaside domain.

\subsection*{Boundary Conditions}
Published literature has demonstrated the improved effects in reducing energy dissipation caused by excessive friction through the use of the new mDBC boundary condition for solving engineering problems whilst using complex geometries with little increase in computational cost \citep{english2019correction, english2022modified, heydemans2024numerical}. It has also been found that the use of mDBC boundary conditions allows for accurate results to be generated from lower resolution setups, reducing runtime \citep{altomare2022latest}. The use of the mDBC boundary is also shown to be more accurate in computing pressure forces on solid surfaces \citep{english2022modified}, making it an ideal solution for tsunami flows and interactions within these cases. As previously mentioned, the use of mDBC boundaries does add increased complexity and computational demand when setting up surface normals. As a result, we use the mDBC boundary when validating against the DBC case with little observed impact. We then use DBC boundary conditions for the larger Cilacap case due to the larger scale of the case and little observed benefit during validation.

Regardless of the advances being made in the use of boundary conditions, there is still the need for the numerical scheme prevent the penetration of particles through the boundary. As per the aforementioned literature, it is observed across these cases that using mDBC boundaries avoid the nonphysical gap between fluid particles and solid particles caused by the repellent forces from changes in pressure and density within the smoothing kernel, and the Case II tuning tests show improved handling of wave reflection and interactions with solid objects such as tank walls, seabeds, and sloped beaches. However, while mDBC greatly reduces numerical instabilities and improves energy conservation near boundaries \citep{english2022modified}, perfect energy conservation is still difficult to achieve in particle-based methods such as SPH, particularly when simulating complex wave simulations \citep{colagrossi2013smoothed}. There is still some degree of numerical friction between fluid and boundary particles, especially in shallow water where wave breaking occurs. Also, as waves approach the shore and the water becomes shallower, their interaction with the boundary becomes more pronounced. Whilst the mDBC provides a more rigorous formulation for the true boundary condition, shallower water scenarios still involve significant numerical diffusion, where small amounts of wave energy are absorbed by the boundary over time. This effect can accumulate, especially with the smoothing effect inherent within SPH, causing noticeable wave energy loss in longer simulations, especially during wave runup or breaking events which are abundant in tsunami simulations \citep{lxxb2022-041}.

\subsection*{Dimensionality}
For Case III, we simulate the case in three dimensions, however it is important to note the differences between simulating in two and three dimensions. When simulating in 2D, the particles interact in a plane, in this case represented by x and z coordinates. The smoothing kernel used in SPH is a function of the distance in two dimensions. The interaction between particles is simpler because it only considers pairwaise interactions in a two-dimensional space, leading to less computational demand and shorter runtimes. In 3D, particles interact in three-dimensional space (x, y, and z). The smoothing kernel must consider distances in three dimensions, making the numerical calculations somewhat more complex. The number of interactions increases rapidly due to the additional spatial dimension. For instance, the number of neighbouring particles that contribute to the smoothing process is larger in 3D, leading to increased computational load and memory requirements. For example, if $dp = 0.5$ (as in Case III), and the smoothing kernel radius, $H$, is 4.15m, a 2D simulation would have 54 particle neighbours within each kernel, whereas a 3D simulation would have 297 particles. In 2D, fluid behaviours such as wave propagation, vortex formation, and interactions tend to be simpler and can be visually represented on a flat surface. In 3D, the same phenomena can exhibit more complex behaviours due to the additional dimension, leading to intricate flow patterns, swirling motions, and mixing processes.

\section{Conclusions}
This study has explored the development and application of the SPH method for simulating tsunami scenarios, emphasising its advantages in handling flows up to and around complex coastal structures and varying topographical features. By validating the SPH approach against two test cases, we demonstrated its capability to generate results that closely align with experimental datasets, thereby confirming its effectiveness as a simulation tool for tsunami flows.

The investigation revealed several critical factors influencing simulation accuracy. The role of artificial viscosity was particularly significant, as it stabilises the numerical computations but can also introduce challenges in accurately capturing wave dynamics. Our findings indicated that while artificial viscosity is essential for mitigating instabilities, its value must be carefully calibrated to avoid excessive energy dissipation. The established value of $\alpha = 0.01$
for wave–structure interactions, while validated, suggests further tuning may enhance simulation fidelity without compromising stability. Particle resolution emerged as another crucial aspect of the simulation development process. The analysis indicated that whilst finer particle spacing can improve the resolution of complex fluid dynamics, it also significantly increases computational costs. The observed particle spacing of $dp=0.05 m$ was demonstrated to be adequate in capturing wave characteristics prompts a reevaluation of conventional guidelines, suggesting that computational efficiency can be achieved without excessively high particle resolutions. This is particularly relevant for large-scale simulations where computational resources are a limiting factor. The use of mDBC proved to be beneficial in reducing numerical friction and improving energy conservation at solid boundaries. By mitigating the non-physical interactions between fluid and boundary particles, mDBC enhances the accuracy of simulations, particularly in shallow water scenarios where wave breaking is prevalent. However, despite these advancements, this development process identified that perfect energy conservation remains difficult to achieve in SPH, especially in scenarios involving complex wave interactions.

As we experimentally transitioned from two-dimensional to three-dimensional simulations, additional complexities highlighted the need for careful parameter selection and geometry setup. The added dimension introduced significant challenges in accurately modeling boundary interactions, particularly in simulating real-world coastal environments. The errors in wave propagation and energy concentration along boundaries highlight the necessity for further refinement in boundary treatment techniques.

This case development is primarily limited in the absence of tuning each and every parameter within DualSPHysics. To allow for maximum applicability and to ensure an almost infinite number of possible cases can be created within DualSPHysics, and the fact that it is designed to be able to simulate a range of flows beyond just water, means there are over 100 possible parameters that could be tested, calibrated, tuned, and validated. Purely on the basis that there is a finite amount of time to progress this research towards the overall goal of emulating these outputs, only a small number of parameters were investigated in this development process. The resources required to finely tune each of these parameters for the purpose of simulating wave-structure interaction with tsunami runup and inundation, constitutes a study in itself, so this is proposed to be one of the next steps in progressing this research.

This study validates the potential of SPH as a viable alternative for tsunami simulations, particularly in scenarios where traditional Eulerian methods may struggle or be insufficient. The results contribute to a deeper understanding of fluid dynamics in tsunami events, with potential future practical implications for tsunami risk assessment and mitigation strategies.

Moving forward, future research should focus on fine-tuning SPH parameters, investigating the impact of alternative viscosity treatments, and expanding the range of test cases to include more complex coastal geometries. Additionally, future releases of software versions that allow parallellisation across multiple GPUs, may permit the scaling of simulations, enabling more realistic and comprehensive modeling of tsunami behaviour.

The next step in this work will take the final case setup outlined in this paper and look to couple this with a NLSWE solver to simulate tsunami flows in deeper water where particle based modelling is unnecessary and computationally inefficient, and then to run this multiple times whilst altering the input parameters to build input-output relationships to then fit emulators to. This will allow us to create an array of predictions that could provide a proof of concept for using SPH simulations in probabilistic tsunami hazard assessments and the scaling of the use of the SPH method in providing meaningful outputs along longer stretches of coastline.


\section*{Acknowledgements}
J.D. was supported by a Studentship from the London Natural
Environment Research Council DTP (grant no. NE/S007229/1). The authors acknowledge the use of the UCL Myriad High Performance Computing Facility (Myriad@UCL), and the Cirrus UK National Tier-2 HPC Service at EPCC (http://www.cirrus.ac.uk) funded by the University of Edinburgh and EPSRC (EP/P020267/1), and associated support services, in the completion of this work.


\newpage
\bibliographystyle{plainnat} 
\bibliography{references.bib} 

\end{document}